# Distribution of the sheet current in a magnetically shielded superconducting filament

S.V. Yampolskii[*], Yu.A. Genenko, H. Rauh

*Institut für Materialwissenschaft, Technische Universität Darmstadt,*

*64287 Darmstadt, Germany*

**Abstract**

The distribution of the transport current in a superconducting filament aligned parallel to the flat surface of a semi-infinite bulk magnet is studied theoretically. An integral equation governing the current distribution in the Meissner state of the filament is derived and solved numerically for various filament-magnet distances and different relative permeabilities. This reveals that the current is depressed on the side of the filament adjacent to the surface of the magnet and enhanced on the averted side. Substantial current redistributions in the filament can already occur for low values of the relative permeability of the magnet, when the distance between the filament and the magnet is short, with evidence of saturation at moderately high values of this quantity, similar to the findings for magnetically shielded strips.

---

[*]On leave from Donetsk Institute for Physics and Technology, National Academy of Sciences of Ukraine, Donetsk 83114, Ukraine.

## 1. Introduction

The effect of strong redistribution of the current in thin superconductor strips or flat superconductor rings caused by magnetic environments has recently been predicted theoretically [1-3] and observed experimentally [4]. Although expected both in the flux-free Meissner state [1] and in the partly flux-filled critical state [2], this effect is most pronounced in the former state where sharp current peaks appear at the edges of the strips [5]; a phenomenon of purely geometrical sort due to the large aspect ratio of thin strips. By attenuating the magnetic field at the edges of the strip, suitably designed magnetic environments facilitate the reduction of such peaks, giving rise to a concomitant increase of the total loss-free current that can be carried by the strips [1,2].

Substantial current enhancements in shielded strips originate from the high sensitivity of the current peaks to the particular form of the magnetic environment, and thus represent a geometrical effect intrinsic to superconductors of a planar shape. Nevertheless, a significant influence of magnetic shielding on critical currents and ac losses has been observed in three-dimensional configurations of composite cables and tapes as well [6-10]. For example, the iron-sheathed superconductor filaments of $MgB_2$/Fe cables exhibit increased critical currents and reduced ac losses in a wide range of the strength of an external magnetic field [8-10]. Cylindrical magnetic shells may, to a certain extent, protect a single filament from the applied field and from the fields due to the presence of other filaments; however, for reasons of symmetry, they cannot protect the filament from the current self-induced magnetic field as in the case of shielded strips [1,2]. Indeed, a cylindrical filament already has the optimum shape in the sense that the current only creates a tangential magnetic field at the surface of the filament; but, taking the fields of the additional filaments into account, asymmetric magnetic shells should be considered too with regard to optimizing the shielding effect. This raises the fundamental point whether such environments are at all capable to strongly influence the current distributions in three-dimensional superconductors where no current peaks occur. We therefore investigate the efficacy of magnetic shielding of a superconducting filament by a bulk magnet, calculating the distribution of the transport current in it for various filament-magnet distances and different relative permeabilities.

## 2. Model

Let us consider a cylindrical superconducting filament of radius $R$, extended infinitely in the $z$ direction of a cartesian coordinate system $x$, $y$, $z$ and located at a distance $a$ from the flat surface of a bulk magnet with relative permeability $\mu$, occupying the space $x \geq a + R$, as depicted in Fig. 1. The filament is supposed to exhibit the flux-free Meissner state, carrying a total current of fixed magnitude, *I*. Provided that the radius of the filament and its distance to the magnet are much larger than the London penetration depth, variations of the magnetic field across the surface layer of the filament may be ignored and, for mathematical convenience, this layer regarded as infinitesimally thin. Such an approach enables the magnetic field $\boldsymbol{H}$ around the filament to be expressed through the sheet current *J* alone which, in cylindrical polar coordinates $(r, \varphi, z)$ adapted to the filament, only depends on $\varphi$. We seek a representation of the latter field in terms of the sheet current as a cylindrical current source, from which a governing equation for *J* results by enforcing the condition of expulsion of magnetic flux typifying the Meissner state.

The magnetic field $\boldsymbol{H}$ is conveniently decomposed according to $\boldsymbol{H} = \boldsymbol{H}^f + \boldsymbol{H}^m$, where $\boldsymbol{H}^f$ stands for the magnetic field created by the shielding current of the filament and $\boldsymbol{H}^m$ denotes the magnetic field induced by the presence of the magnet adjacent to the filament. Together with the concept of line current sources [11], distributed continuously along the circumference of the filament, and the concept of equidirectional line current images [12], distributed continuously along the circumference of the image of the filament regarding the mirror plane $x = a$, this ansatz ensures continuity of the tangential component of the magnetic field and of the normal component of the magnetic induction when the surface of the magnet is traversed. Invoking Ampère's law applied to line current sources with radius vector $\boldsymbol{r}^s$ (see Fig. 1), the radial and azimuthal components of $\boldsymbol{H}^f$ in $r \geq R$, $0 \leq \varphi < 2\pi$ read:

$$H_r^f(r,\varphi) = \frac{1}{2\pi} \int_0^{2\pi} d\varphi' J(\varphi') \frac{R^2 \sin(\varphi-\varphi')}{r^2 + R^2 - 2rR\cos(\varphi-\varphi')},$$

$$H_\varphi^f(r,\varphi) = \frac{1}{2\pi} \int_0^{2\pi} d\varphi' J(\varphi') \frac{R(r-R\cos(\varphi-\varphi'))}{r^2 + R^2 - 2rR\cos(\varphi-\varphi')}. \quad (1)$$

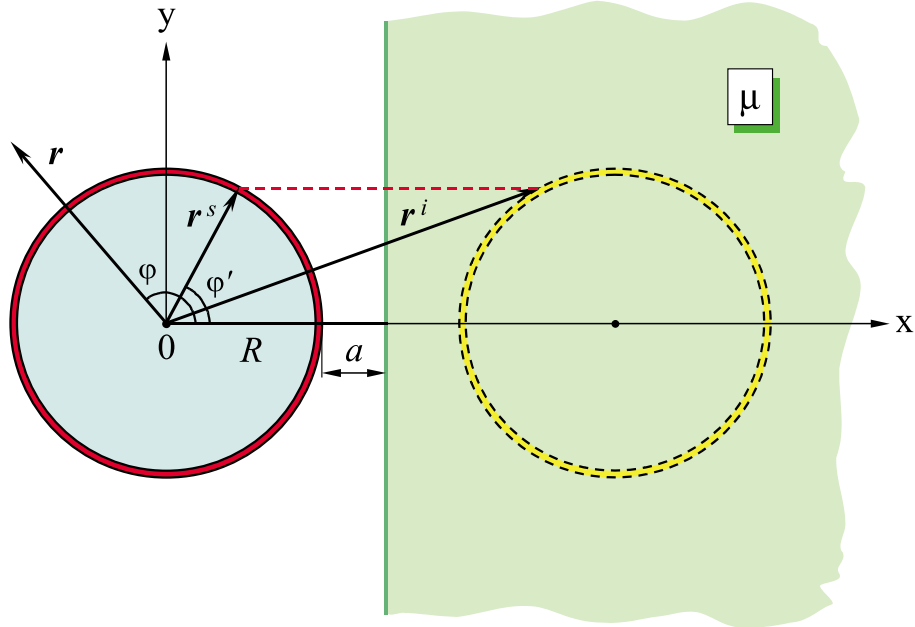

Fig. 1. Cross-sectional view of the superconducting filament of radius *R* located at a distance *a* from the flat surface of the bulk magnet (green shading). The red ring indicates the current-carrying surface layer of the filament (thickness not to scale!), $\boldsymbol{r}^s$ representing the radius vector of a line current source; the yellow ring indicates the image of this layer with respect to the surface of the magnet, $\boldsymbol{r}^i$ representing the radius vector of a line current image. Cylindrical polar coordinates $(r, \varphi, z)$ of a line outside the filament with radius vector $\boldsymbol{r}$ are marked.

Furthermore, from the line current images with radius vector $\boldsymbol{r}^i$ (see Fig. 1), the radial and azimuthal components of $\boldsymbol{H}^m$ in $r \geq R$, $0 \leq \varphi < 2\pi$ are:

$$H_r^m(r,\varphi) = \frac{q}{2\pi} \int_0^{2\pi} d\varphi' J(\varphi') \frac{R(R\sin(\varphi+\varphi') - 2(a+R)\sin\varphi)}{(r\cos\varphi + R\cos\varphi' - 2(a+R))^2 + (r\sin\varphi - R\sin\varphi')^2},$$

$$H_\varphi^m(r,\varphi) = \frac{q}{2\pi} \int_0^{2\pi} d\varphi' J(\varphi') \frac{R(r+R\cos(\varphi+\varphi') - 2(a+R)\cos\varphi)}{(r\cos\varphi + R\cos\varphi' - 2(a+R))^2 + (r\sin\varphi - R\sin\varphi')^2}, \quad (2)$$

where $q = (\mu-1)/(\mu+1)$ reflects the image current strength.

Expulsion of magnetic flux in the Meissner state is guaranteed, according to the supposition of an infinitesimally thin current-carrying surface layer stated before, by the vanishing of the radial component of the magnetic field on the surface of the filament, $H_r = 0$ for $r = R$. From the decomposition of the magnetic field in conjunction with Eqs. (1) and (2), this requirement yields the Fredholm integral equation of the first kind for the cylindrical sheet current in $0 \leq \varphi < 2\pi$:

$$\int_0^{2\pi} d\varphi' J(\varphi') M(\varphi,\varphi') = 0 \quad (3)$$

with the singular kernel

$$M(\varphi,\varphi') = \cot\left(\frac{\varphi-\varphi'}{2}\right) - q \frac{\sin(\varphi+\varphi') - 2(1+\alpha)\sin\varphi}{\cos(\varphi+\varphi') - 2(1+\alpha)(\cos\varphi + \cos\varphi') + 2(1+\alpha)^2 + 1}, \quad (4)$$

the ratio $\alpha = a/R$ herein measuring the distance between the filament and the magnet relative to the radius of the filament. By employing Hilbert's identity [13] in $0 \leq \varphi < 2\pi$,

$$\frac{1}{4\pi^2}\int_0^{2\pi} d\varphi' \cot\left(\frac{\varphi-\varphi'}{2}\right)\int_0^{2\pi} d\varphi'' J(\varphi'')\cot\left(\frac{\varphi'-\varphi''}{2}\right) + J(\varphi) = \frac{1}{2\pi}\int_0^{2\pi} d\varphi' J(\varphi'), \quad (5)$$

the singularity of Eq. (3) is aptly removed and Eq. (3) itself transformed into the equivalent Fredholm integral equation of the second kind for the cylindrical sheet current in $0 \leq \varphi < 2\pi$,

$$\int_0^{2\pi} d\varphi' J(\varphi') K(\varphi,\varphi') + J(\varphi) = (1+q)\frac{I}{2\pi R} \quad (6)$$

with the regular kernel

$$K(\varphi,\varphi') = \frac{q}{\pi}\frac{(1+\alpha)(1+\alpha-\cos\varphi')}{\cos(\varphi+\varphi') - 2(1+\alpha)(\cos\varphi+\cos\varphi') + 2\alpha(2+\alpha) + 3}, \quad (7)$$

the total current

$$I = R\int_0^{2\pi} d\varphi' J(\varphi'), \quad (8)$$

by definition of the problem, being fixed. Equation (6) is readily amenable to efficient numerical investigations using standard computational schemes [14].

## 3. Results and discussion

To appraise the effect of magnetic shielding on the current flow in the superconducting filament, Eq. (6) was solved numerically for a range of the geometrical and material parameters involved, and the results were displayed graphically. Figure 2 shows the angular variation of the sheet current for different values of the distance between the filament and the magnet, when the value of the relative permeability of the magnet is prescribed. A general trait due to current redistribution in the filament is the depression of the sheet current on the side of the filament adjacent to the surface of the magnet and an enhancement with a wide maximum on the averted side. Whereas this effect is distinctly more pronounced for the higher of the two values of the relative permeability singled out, including, at short filament-magnet distances, an almost complete suppression of the sheet current on the adjacent side, it abates when the filament-magnet distance is increased, with a tendency to reducing the sheet current anisotropy. The angular variation of the sheet current for different values of the relative permeability of the magnet, with the distance between the filament and the magnet prescribed, is portrayed in Fig. 3. This reveals that, whereas the effect of current redistribution is distinctly more pronounced for the lower of the two values of the filament-magnet distance set, including an almost complete suppression of the sheet current on the adjacent side and saturation for high values of the relative permeability, it abates again when the filament-magnet distance is increased, with a tendency to mitigating the sheet current anisotropy. The asymptotic behaviour with respect to the relative permeability is evident from Eq. (6).

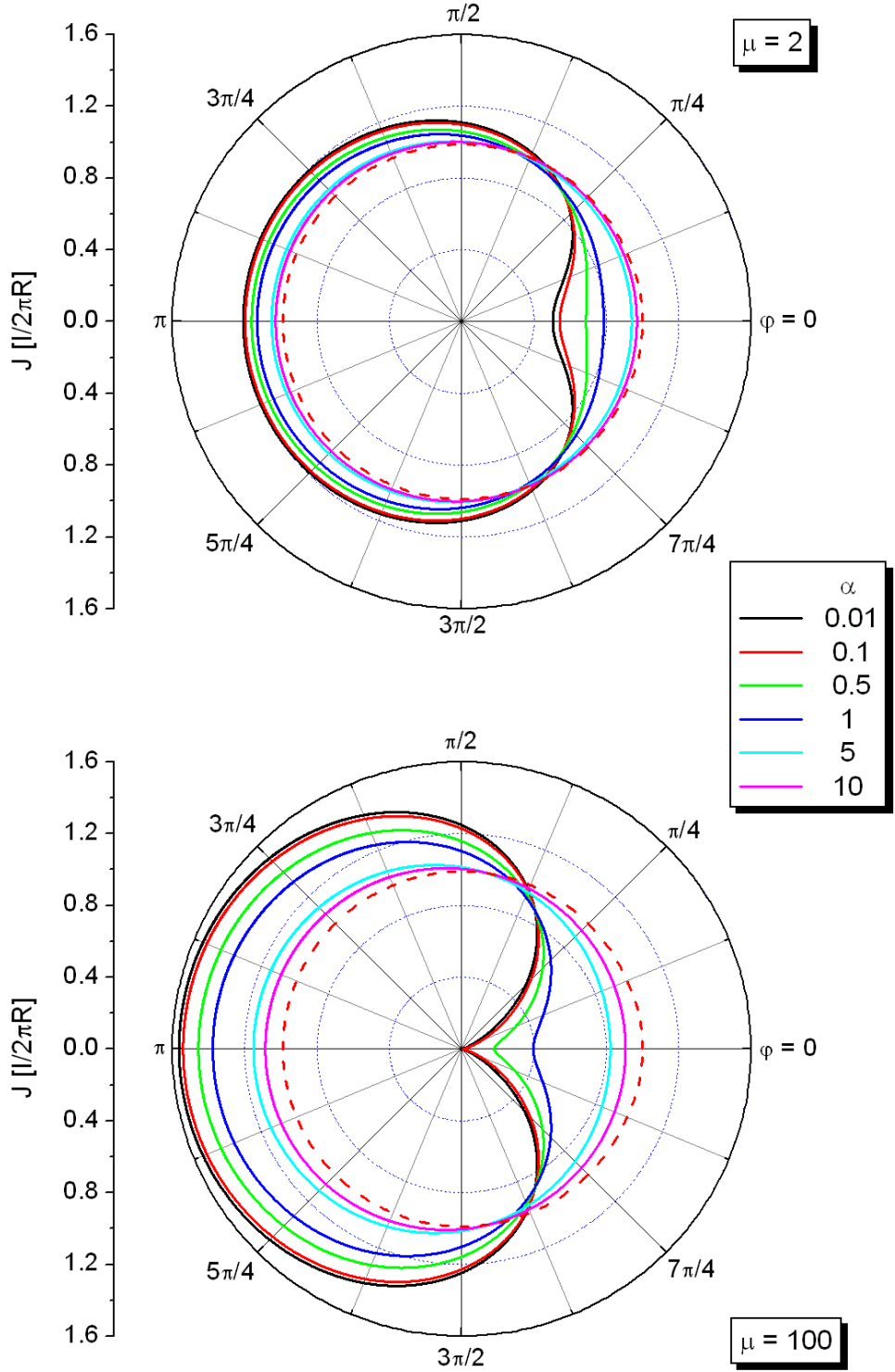

Fig. 2. Distribution of the sheet current of the superconducting filament, $J$, as a function of the azimuth, $\varphi$, for different values of the distance between the filament and the magnet relative to the radius of the filament, $\alpha$, specified in the legend, when the value of the relative permeability of the magnet, $\mu = 2$ (upper part) and $\mu = 100$ (lower part). The isotropic distribution of the sheet current in the absence of the magnet (i.e. for $\mu = 1$), represented by the unit circle (red dashed line), is shown for comparison.

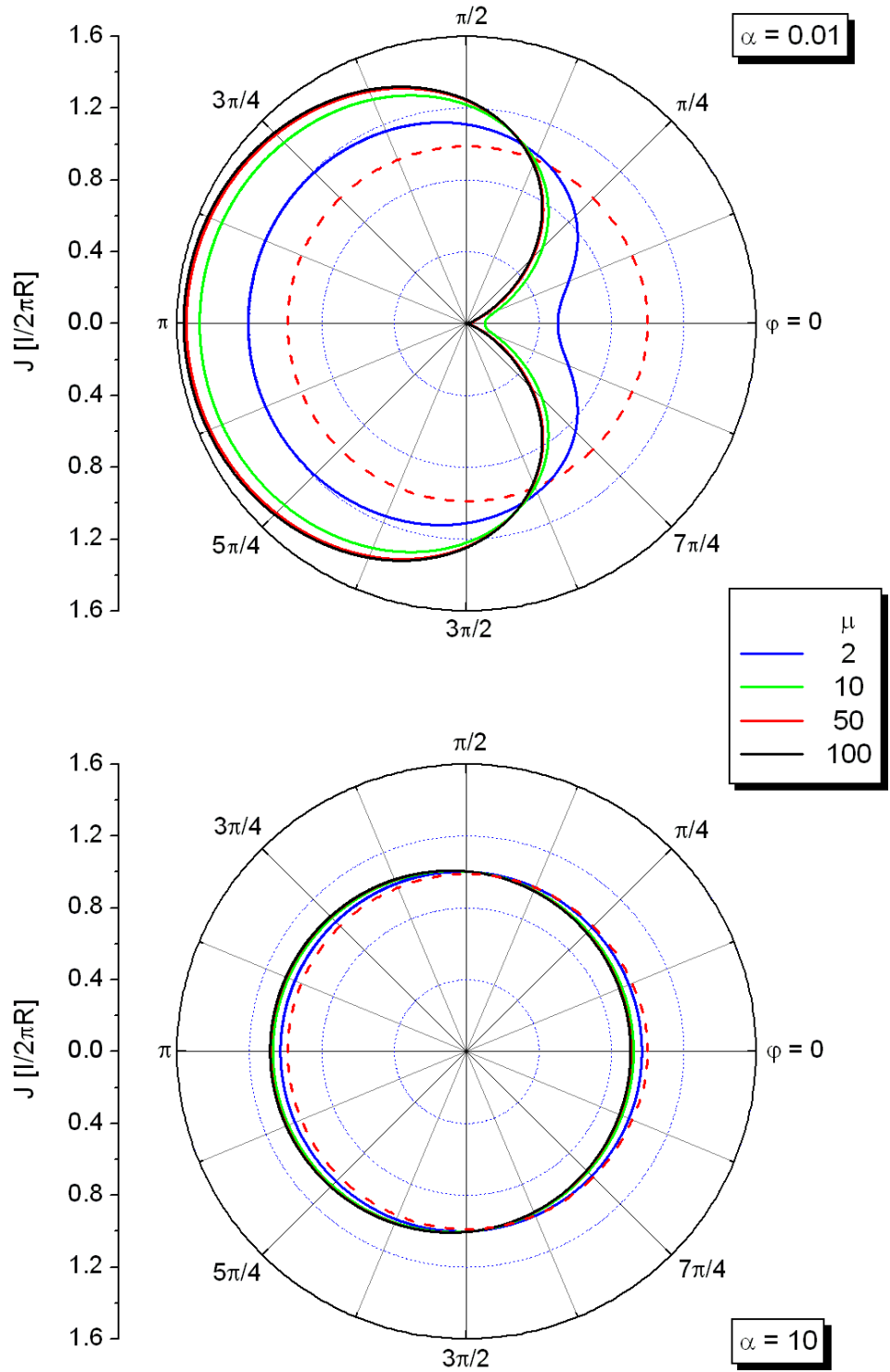

Fig. 3. Distribution of the sheet current of the superconducting filament, $J$, as a function of the azimuth, $\varphi$, for different values of the relative permeability, $\mu$, specified in the legend, when the value of the distance between the filament and the magnet relative to the radius of the filament, $\alpha = 0.01$ (upper part) and $\alpha = 10$ (lower part). The isotropic distribution of the sheet current in the absence of the magnet (i.e. for $\mu = 1$), represented by the unit circle (red dashed line), is shown for comparison.

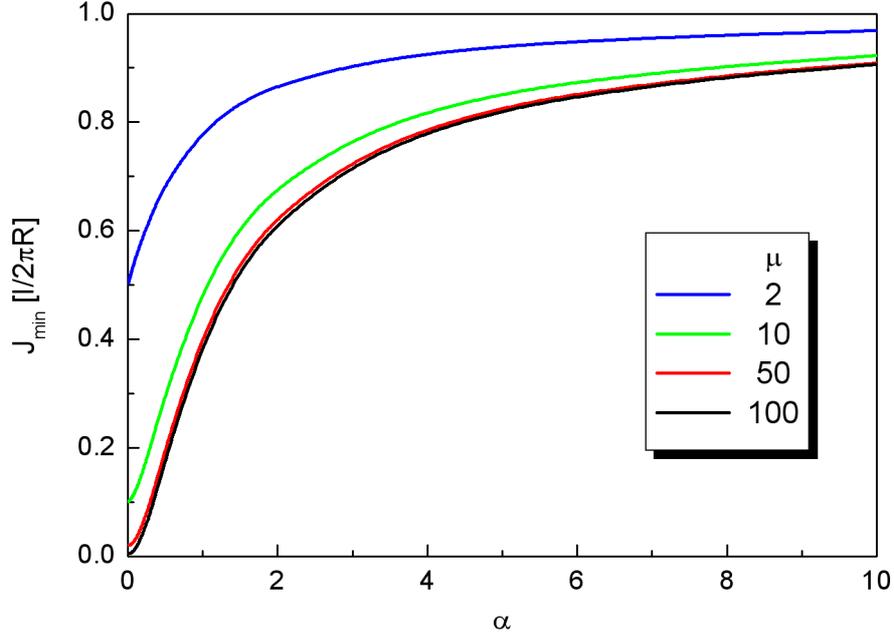

Fig. 4. Minimum value of the sheet current of the superconducting filament, $J_{min}$, as a function of the distance between the filament and the magnet relative to the radius of the filament, $\alpha$, for different values of the relative permeability, $\mu$, specified in the legend. The minimum value of the sheet current in the absence of the magnet (i.e. for $\mu = 1$), adopted independently of $\alpha$, is unity.

Figure 4 shows the variation of the minimum value of the sheet current at zero azimuth with the distance between the filament and the magnet relative to the radius of the filament, when the value of the relative permeability of the magnet is prescribed. A general trait due to current redistribution in the filament is the increase of the minimum value of the sheet current with the filament-magnet distance, but a decrease with the relative permeability, the effect saturating for large values of either of these quantities. In the limit of zero filament-magnet distance, the minimum value adopts a finite magnitude and finite derivative, unlike in the case of shielded strips, where divergent derivatives persist [1,2]. The variation of the minimum value of the sheet current at zero azimuth with the relative permeability, for different values of the distance between the filament and the magnet relative to the radius of the filament prescribed, is portrayed in Fig. 5. This reveals that the minimum value of the sheet current decreases with the relative permeability due to current redistribution in the filament, but increases with the filament-magnet distance, the effect saturating again for large values of either of these quantities. As before, the asymptotic behaviour with respect to the relative permeability is evident from Eq. (6).

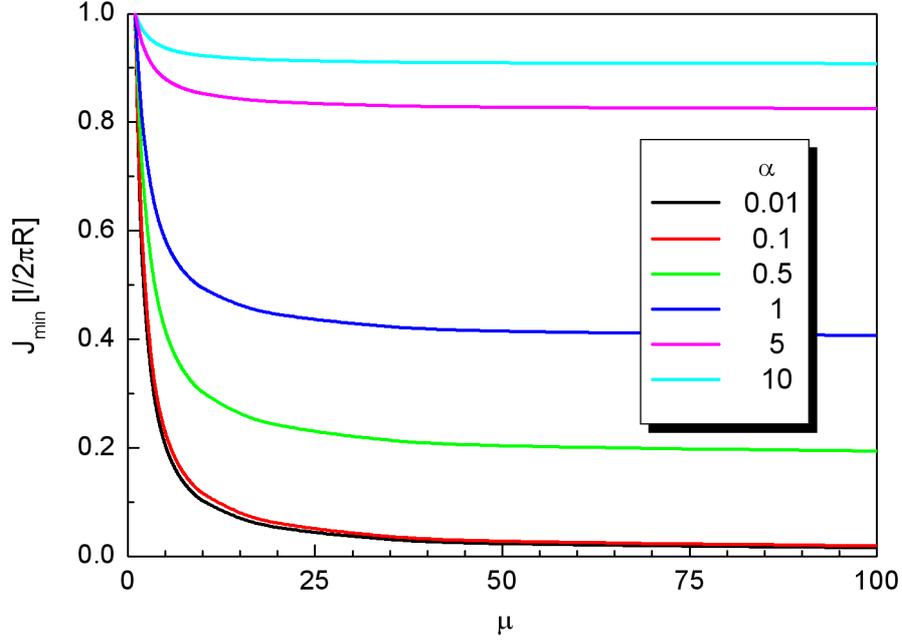

Fig. 5. Minimum value of the sheet current of the superconducting filament, $J_{min}$, as a function of the relative permeability, $\mu$, for different values of the distance between the filament and the magnet relative to the radius of the filament, $\alpha$, specified in the legend. The minimum value of the sheet current in the absence of the magnet (i.e. for $\mu = 1$), adopted independently of $\alpha$, is unity.

We comment that the phenomenon of depression of the sheet current on the side of the filament adjacent to the surface of the magnet, with a minimum value at zero azimuth, can easily be understood by recalling the current distribution observed in thin superconductor strips. Isolated strips exhibiting the Meissner state are known to show a transport current distribution with sharp peaks at the edges and a minimum in the centre of the strips [5,15,16]. This type of distribution is impelled by the requirement of a zero normal component of the magnetic field on the surface of the strips in the Meissner state; a constraint preponderating over the effect of mutual attraction of equidirectional line current sources, distributed continuously across the strips, which would create minima at the edges and a maximum in the centre of the strips. If a current-carrying strip were conceived as cut into two parallel half-strips joined together along their common edge, the distribution of the sheet current in these half-strips thus would show peaks at their respective outer edges, but a depression with a minimum at the contact edge. Precisely the same view also applies to the configuration of a current-carrying half-strip located near - and oriented orthogonal to - the flat surface of a semi-infinite bulk magnet, which generates an image of the half-strip carrying a current of the original sign. In the limit of direct contact between the half-strip and the surface of the magnet, the half-strip together with its image effectively form a complete strip of full width, which reveals a minimum of the sheet current in the centre of the strip, i.e. at the physical edge of the half-strip adjacent to the surface of the magnet. Quantitative investigations of the latter configuration unequivocally substantiate the correctness of this result [2].

To conclude, the distribution of the transport current in a cylindrical flux-free filament aligned parallel to the flat surface of a semi-infinite bulk magnet was investigated both analytically and numerically. The results demonstrate a dependence of the sheet current on the filament-magnet distance very much like for the case of magnetically shielded strips [1,2], though without current peaks typical of strips [5]. Substantial current redistributions in the filament can already occur for low values of the relative permeability of the magnet, when the distance between the filament and the magnet is short, with evidence of saturation at moderately high values of this quantity, similar to the findings for magnetically shielded strips [1,2]. We therefore expect that an asymmetric magnetic shielding configuration is indeed capable to redistribute the current in a superconducting filament efficaciously.

**Acknowledgment**

This work was supported by a grant from the German Research Foundation (DFG).